# AQR-HNSW: Accelerating Approximate Nearest Neighbor Search via Density-aware Quantization and Multi-stage Re-ranking

Ganap Ashit Tewary
gtewary@asu.edu
Arizona State University
Tempe, AZ, USA

Nrusinga Charan Gantayat
ngantaya@asu.edu
Arizona State University
Tempe, AZ, USA

Jeff Zhang
jeffzhang@asu.edu
Arizona State University
Tempe, AZ, USA

## Abstract

Approximate Nearest Neighbor (ANN) search has become fundamental to modern AI infrastructure, powering recommendation systems, search engines, and large language models across industry leaders from Google to OpenAI. Hierarchical Navigable Small World (HNSW) graphs have emerged as the dominant ANN algorithm, widely adopted in production systems due to their superior recall vs. latency balance. However, as vector databases scale to billions of embeddings, HNSW faces critical bottlenecks: memory consumption expands, distance computation overhead dominates query latency, and it suffers suboptimal performance on heterogeneous data distributions. This paper presents Adaptive Quantization and Rerank HNSW (AQR-HNSW), a novel framework that synergistically integrates three strategies to enhance HNSW's scalability. AQR-HNSW introduces (1) density-aware adaptive quantization, achieving 4× compression while preserving distance relationships; (2) multi-state re-ranking that reduces unnecessary computations by 35%; and (3) quantization-optimized SIMD implementations delivering 16-64 operations per cycle across architectures. Evaluation on standard benchmarks demonstrates 2.5-3.3× higher queries per second (QPS) than state-of-the-art HNSW implementations while maintaining 98%+ recall, with 75% memory reduction for the index graph and 5× faster index construction.

## 1 Introduction

The exponential growth of high-dimensional data has fundamentally transformed computational challenges in domains spanning machine learning, computer vision, natural language processing, and information retrieval. Modern applications routinely process billion-scale datasets with embedding representations ranging from 128 to 4096 dimensions [8, 13, 25, 34, 36] where the traditional *exact* nearest neighbor search becomes computationally prohibitive on high-dimensional queries. This computational bottleneck has driven widespread adoption of Approximate Nearest Neighbor (ANN) algorithms, which deliver orders-of-magnitude performance improvements with marginal accuracy degradation, enabling real-time similarity search at scale [21].

Among ANN approaches, graph-based [6, 15, 35] approaches have emerged as the dominant paradigm due to their superior recall-latency characteristics compared to tree-based [4, 26, 28] and quantization-based methods [3, 16, 24]. Hierarchical Navigable Small World (HNSW) graphs, introduced by Malkov and Yashunin, designed for high-dimensional similarity search, demonstrating consistent superiority across diverse datasets and query distributions [23]. HNSW's multi-layer probabilistic skip-list structure enables logarithmic search complexity while maintaining near-optimal recall through navigation across hierarchical graph layers.

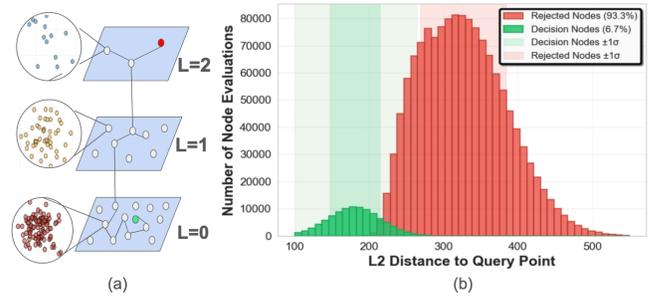

Figure 1: (a) HNSW navigates from sparse to dense nodes to reach the nearest neighbor. (b) Distribution of node evaluations during HNSW search.

The adoption of HNSW has accelerated across both academic and industrial domains. Production systems such as Meta [19] and Spotify [22] rely on HNSW-based implementations for datacenter applications, e.g., personalized recommendation engines serving billions of users, retrieval-augmented generation (RAG) systems powering large language models, content-based search across multimedia databases. Open-source vector databases such as ElasticSearch [12], Pinecone [30], Weaviate [38], and Qdrant [33] have standardized HNSW as their core indexing mechanism. However, HNSW still requires trillions of operations for production systems.

HNSW's hierarchical structure exhibits density variations across layers shown in Figure 1(a) on various datasets. The algorithm begins search at sparse upper layers (L=2, shown in blue) where data points are sparsely distributed, then progressively descends through increasingly dense middle layers (L=1, orange nodes) before reaching the densest base layer (L=0, red nodes) where the final nearest neighbors reside. This traversal requires evaluating numerous candidate nodes at each layer, computing distances to determine which paths lead closer to the query point.

However, Figure 1(b) showcase a critical inefficiency in current HNSW implementations: while 93% of evaluated nodes are ultimately rejected during the search process. Traditional approaches compute full FP32 precision distances for all nodes uniformly which creates a substantial wasted distance computation. Only a small fraction (7%) of evaluated nodes, as referred to as "decision nodes," actually influence the final search, but are treated equally in terms of their precision as all evaluated nodes.

To accelerate HNSW while maintaining high-quality search, we leverage three key observations. First, the vast majority of rejected nodes present an opportunity for aggressive computational savings through reduced-precision approximations, as these nodes only need to be identified as non-competitive rather than ranked precisely. Second, the density stratification visible in Figure 1(a) enables adaptive precision allocation: sparse upper layers with widely separated nodes can tolerate coarse distance approximations, while dense lower layers with tightly clustered neighbors require higher precision to distinguish between competitive candidates. Third, the hierarchical navigation naturally supports progressive refinement, where approximate distances guide early layer traversal efficiently, while exact computations are reserved for the critical later-layer decision nodes that determine final accuracy.

To this end, we present Adaptive Quantized and Reranking HNSW (AQR-HNSW), a novel approach that fundamentally rethinks how distance computations are performed in graph-based ANN search. Our key insights are that different regions of the vector space and different stages of the search process require different levels of precision. The main contributions of this work are:

(1) **Density-aware Adaptive Quantization:** A novel quantization scheme that dynamically adjusts precision based on data density, i.e., allocating more bits to dense regions while aggressively compressing sparse regions.

(2) **Recall-aware Adaptive Re-ranking:** An multi-stage re-ranking mechanism that progressively refines search using fixed-point format and performs expensive FP32 distance computations only when necessary to maintain recall level.

(3) **Comprehensive Empirical Validation:** Across standard benchmarks, AQR-HNSW achieves 1.8–2.5× higher throughput and faster query-time latency with marginal recall loss. The index build time is reduced by 5× via our density-aware quantization.

The remainder of this paper is organized with Section II provides background on HNSW. Section III presents our AQR-HNSW methodology. Section IV presents comprehensive experimental evaluation. Finally, Section V concludes the work.

## 2 Background and Motivation

HNSW algorithm constructs a hierarchical structure of a proximity graph by assigning each graph element to multiple layers with exponentially decreasing probability. The algorithm maintains three critical parameters: **M** (the maximum number of connections per node), **ef_construction** (the size of the dynamic candidate list during the graph construction) and **ef_search** (the size of the dynamic candidate list during search). Navigation starts from an entry point in the highest layer and greedily traverses to lower layers, progressively searching until the nearest-neighbors are found [23].

Industrial deployments have driven significant HNSW enhancements. Spotify's Voyager [22] serves 1+ billion music embeddings with memory-mapped indices and SIMD-accelerated systems, achieving 1M QPS in production. Elasticsearch's [12] implementation adds real-time updates at 100K documents/second through concurrent graph construction. Academic advances include NSW-Flash [7] for SSD-optimized search, FilteredHNSW [31] for metadata-aware retrieval achieving 95% filtered recall, and LearnedHNSW [9] using GNNs to optimize graph topology. Despite these improvements, fundamental bottlenecks persist: distance computations consume 70-85% of search time, and static hyper-parameters perform poorly on heterogeneous data vector distribution.

**Efficient ANN search and Their Limitations** Several works have explored efficient techniques to optimize the ANN search as standalone improvement, however, their synergistic integration with HNSW algorithms are significantly underexplored.

**Quantization** Quantization techniques have become central to scaling vector search, with scalar and binary methods offering lightweight compression and product-based approaches enabling deeper optimization. Scalar quantization reduces floating-point precision to fixed-point formats, with Int8 variants achieving 4× compression and 95% recall, as adopted in Pinecone [30] and Weaviate [38] binary quantization. FAISS [19] compresses vectors to single bit, yielding 32× compression but with a recall degradation to 75–80% in high-dimensional spaces.

**Dynamic Parameter Adjustment.** Dynamic HNSW (DHNSW) [18] adjusts M and ef_construction based on local density. However, DHNSW only optimizes graph construction, not query-time performance. Adaptive beam search in DiskANN [15] dynamically adjusts search width based on intermediate recall estimates, improving 99-percentile latency by 2×. Learning-based approaches like NeuralLSH [32] and LearnedIndex [10] predict optimal parameters using query features but require extensive training data.

**Early Termination.** Early termination strategies such as NSW-Flash [6] using local minimum detection, IVF [19] employing cell-level pruning, and LSH [20] applying progressive candidate filtering. The challenge is that terminating too early sacrifices recall but conservative termination provide minimal speedup. Existing methods either require training on query logs or use fixed thresholds that are unsuitable for heterogeneous data.

**Motivation for AQR-HNSW.** (1) Current methods optimize either memory (through quantization) or computation (through parameter tuning) only. Existing quantization reduces storage but distance computations remain expensive; dynamic parameter adjustment reduces build time but not query-time latency. AQR-HNSW addresses both through adaptive quantization and multi-stage re-ranking. (2) Existing techniques overlook non-uniform data distribution and assume static performance requirements. Real production workloads exhibits density variations and varying recall needs (e.g., 95% for recommendation [5, 11, 27], 99.9% for compliance[11, 21]). AQR-HNSW adapts to these: quantization precision, re-ranking depth, early termination thresholds, providing adaptive performance-recall tradeoffs across diverse use-cases.

## 3 AQR-HNSW Framework

This section introduces AQR-HNSW, a comprehensive framework that achieves higher QPS while maintaining search quality through density-aware quantization and multi-stage adaptive retrieval. The complete workflow comprising two phases: the **Build Phase** as described in Algorithm 1, which processes dataset insertion through density-aware 8-bit quantization, and the **Search Phase** as described in Algorithm 2, which performs coarse-to-exact retrieval with early termination.

### 3.1 Build Phase: Index Construction

When a dataset $\mathcal{D} = \{\mathbf{x}_1, \ldots, \mathbf{x}_n\}$ is inserted into AQR-HNSW, it undergoes a three-stage transformation process that constructs a quantized graph index optimized for search performance.

**Stage 1: Density Characterization**

Upon receiving the dataset, the system first analyzes the local data distribution to identify regions of varying point concentration. For each data point $\mathbf{x}_i \in \mathcal{D}$, we compute its local density using $k$-nearest neighbor analysis by $\rho_i = \frac{1}{\bar{d}_k(\mathbf{x}_i)+\epsilon}$ where the average distance to $k$ nearest neighbors (typically $k = 10$) is $\bar{d}_k(\mathbf{x}_i) = \frac{1}{k}\sum_{\mathbf{x}_j \in \text{kNN}(\mathbf{x}_i,k)} \|\mathbf{x}_i - \mathbf{x}_j\|^2$ with $\epsilon = 10^{-6}$ providing numerical stability mentioned in Algorithm 1 (Line 5).

**Efficient computation for large datasets.** When $\mathbf{x}_i > 10{,}000$, we compute densities for a sample of $n_s = \min(5000, n)$ points and assign the average sample density $\bar{\rho} = \frac{1}{n_s}\sum_{i=1}^{n_s} \rho_i$ to the remaining points. The global heterogeneity measure $\delta \in [0, 1]$ quantifies density variation across the dataset defined in Algorithm 1 (Line 6). Values approaching 1 indicate highly non-uniform distributions with distinct dense and sparse regions.

**Stage 2: Adaptive Quantization Training**

After characterizing the data distribution, the system learns dimension-specific quantization parameters that preserve distance relationships while achieving 4× compression.

**Dimension-wise Range Determination.** For each dimension $j \in \{1, \ldots, d\}$, we determine quantization boundaries using adaptive percentiles. This ensures that the encoding process dynamically balances coverage of the full distribution with robustness against outliers. The percentile bounds $[P_l, P_h]$ adjust based on the density variation factor $\delta$ with the relation of $P_l = \delta \cdot P_{\max}$ and $P_h = 1 - \delta \cdot P_{\max}$ where $P_{\max}$ defines the maximum percentile width (sensitivity range) which are used in Algorithm 1 (line 8). The bounds are constrained such that $0 \leq P_l \leq \delta \cdot P_{\max}$, $1-\delta \cdot P_{\max} \leq P_h \leq 100$. For each dimension $j$, we extract values across all data points and compute the lower bound $\min_j$ by taking the $P_l$ percentile and upper bound $\max_j$ by taking the $P_h$ percentile. The scaling factor that maps this range to 8-bit representation is given by $\text{scale}_j = \frac{255}{\max_j - \min_j + \epsilon}$ used as value in Algorithm 1 (Line 11).

**Vector Encoding.** Each incoming vector $\mathbf{x} \in \mathbb{R}^d$ is transformed into its quantized representation, which is element-wise shifted and scaled into the 0–255 range, clipped to stay within the percentile bound $[P_l, P_h]$ range, and then rounded to the nearest integer to get the vectors quantized form as given in Algorithm 1 (Line 9).

**Distance Preservation.** To maintain proportional distances after quantization, we compute a global scaling which is used for minimize the density error due to quantization. First, we calculate per-dimension density weights given in Algorithm 1 (Line 12). Then the distance scale factor is derived as:

$$s_{\text{dist}} = \sqrt{\frac{\sum_{j=1}^{d}(\max_j - \min_j)^2 \cdot w_j}{\sum_{j=1}^{d} 255^2 \cdot w_j}} \quad (1)$$

**Dimension-wise Density Heterogeneity.** To quantify how density varies across dimensions, we compute the coefficient of variation of per-dimension density weights where $\sigma_w = \sqrt{\frac{1}{d}\sum_{j=1}^{d}(w_j - \bar{w})^2}$ is the standard deviation of density weights $\{w_1, \ldots, w_d\}$ from Algorithm 1 (Line 12), and $\bar{w} = \frac{1}{d}\sum_{j=1}^{d} w_j$ is their mean. The ratio of $\sigma_w$ to $\bar{w}$ results in the value of $\eta$ which is defined in Algorithm 1 (Line 14). High value of $\eta$ indicates that density is unevenly distributed across dimensions.

**Connectivity Adjustment.** The number of connections per node is increased when both overall density variation and dimensional heterogeneity are high and the $M_i$ is adjusted and given in Algorithm 1 (Line 17).

**Construction Search Depth.** The search depth during construction is inversely adjusted based on the expected connectivity increase in base search depth and given in Algorithm 1 (Line 18).

---

**Algorithm 1** AQR-HNSW: Dataset Insertion and Graph Construction

1: **Input:** Dataset $\mathcal{D} = \{\mathbf{x}_1, \ldots, \mathbf{x}_n\} \subset \mathbb{R}^d$, $M_0$, $ef_0^{\text{construction}}$
2: **Output:** AQR-HNSW index $\mathcal{G}$
3:
4: // *Stage 1: Density characterization*
5: Compute $\{\rho_1, \ldots, \rho_n\}$ using $k$-NN
6: Compute $\delta = \frac{\max_i(\rho_i) - \min_i(\rho_i)}{\max_i(\rho_i)+\epsilon}$
7: // *Stage 2: Quantization training*
8: Determine $P_l = \delta \cdot P_{\max}$ and $P_h = 1 - \delta \cdot P_{\max}$
9: $\mathbf{q}_i \leftarrow \text{Quantize}(\mathbf{x}_i)$
10: **for** $j = 1$ **to** $d$ **do**
11:     Compute $\min_j, \max_j, \text{scale}_j$
12:     Compute $w_j = \frac{1}{n}\sum_{i=1}^{n} \rho_i$
13: **end for**
14: Compute $s_{\text{dist}}, \eta = \frac{\sigma_w}{\bar{w}}$     ▷ Eq. 1
15: // *Stage 3: Graph construction*
16: $\mathcal{G} \leftarrow \text{InitializeHNSW}()$
17: Compute $M_i = \lfloor M_0 \cdot (1 + \delta \cdot (1 + \eta)) \rfloor$
18: Compute $ef^{\text{construction}} = \left\lfloor ef_0^{\text{construction}} \cdot \frac{1}{1+\delta \cdot \eta} \right\rfloor$
19: **for** $i = 1$ **to** $n$ **do**
20:     $\mathcal{G}.\text{Insert}(\mathbf{q}_i, i, M_i, ef^{\text{construction}})$
21:     Store $\mathbf{x}_i$ for exact reranking
22: **end for**
23: **return** $\mathcal{G}$

---

## 3.2 Search Phase: Progressive Refinement

When a query $\mathbf{q}$ arrives, AQR-HNSW performs approximate $k$-NN retrieval through progressive refinement across three stages. The query is first quantized using learned parameters from the Build Phase.

**Stage 1: Coarse Quantized Search**

The initial retrieval operates entirely in the compressed 8-bit space. Using the quantized query $\hat{\mathbf{q}}$, the distances are computed using 8-bit quantized vectors $\hat{\mathbf{x}}_n$ from the relation $dist_{\text{quantized}}(\hat{\mathbf{q}}, \hat{\mathbf{x}})$ given in Algorithm 2 (line 6). The graph traversal retrieves the top $N_c$ coarse candidates by exploring the HNSW structure with search depth $ef_{\text{search}} = N_c \times m_{ef}$, where $m_{ef}$ is a multiplier that controls the breadth of exploration. This produces a candidate set which has the information of the vector and the distance difference with respect to the query $\hat{\mathbf{x}}$ in $C = \{(\hat{\mathbf{x}}_1, d_1^q), \ldots, (\hat{\mathbf{x}}_{N_c}, d_{N_c}^q)\}$ sorted in ascending order bases of the distance difference.

**Stage 2: Asymmetric Distance Refinement**

To improve ranking accuracy without accessing the full-precision vectors, we employ asymmetric distance computation. Each quantized candidate $\hat{\mathbf{x}}_i$ is decoded to its approximate original representation by $\tilde{x}^{(j)} = \frac{\hat{x}^{(j)}}{\text{scale}_j} + \min_j$ producing reconstructed vector $\tilde{\mathbf{x}} \in \mathbb{R}^d$. The asymmetric distance then compares the full-precision query against this reconstruction given in Algorithm 2 (line 9). This yields refined candidate ranking in the list $\mathcal{A} = \{(\tilde{\mathbf{x}}_1, d_1^a), \ldots, (\tilde{\mathbf{x}}_{N_c}, d_{N_c}^a)\}$ in ascending order based on the asymmetric distance difference.

**Early Termination.** Before committing to expensive exact distance computations, the system analyzes the separation between the top-$k$ candidates and the remaining pool. When $|\mathcal{A}| > k$, we compute two separation metrics: gap = $\frac{d_{k+1}^a - d_k^a}{d_k^a + \epsilon}$ and ratio = $\frac{d_{k+1}^a}{d_k^a + \epsilon}$ where $d_k^a$ denotes the asymmetric distance to the $k$-th nearest candidate. These metrics quantify both the relative gap and the proportional separation at the decision boundary.

If either metric exceeds its corresponding threshold (gap > $\tau_{\text{gap}}$ or ratio > $\tau_{\text{ratio}}$), the system concludes that the top-$k$ results are clearly separated from remaining candidates given in Algorithm 2 (line 16). In this case, exact reranking is limited to at most $k$ candidates, as further refinement is unlikely to alter the top-$k$ ranking.

**Stage 3: Exact Distance Computation**

When early termination is not triggered, indicating ambiguous ranking among candidates, the system evaluates the spread of asymmetric distances. For the selected $N_{\text{rerank}}$ candidates, the system retrieves their original full-precision vectors from storage and computes exact Euclidean distances given in Algorithm 2 (line 19). This produces the exact-distance list $\mathcal{R}_{\text{exact}} = \{(x_i, d_i^e)\}_{i=1}^{N_{\text{rerank}}}$.

**Final Result Assembly** The final result set $\mathcal{R}$ is formed by combining the exact-distance candidates with the remaining asymmetric-distance candidates from $\mathcal{A}$ as needed to ensure at least $k$ results. After sorting by distance in ascending order, the top-$k$ neighbors are returned as the final query result mentioned in Algorithm 2 (Line 23-24).

**Algorithm 2** AQR-HNSW: Query Processing with Progressive Refinement

1: **Input:** Query $\mathbf{q} \in \mathbb{R}^d$, index $\mathcal{G}$, $k$
2: **Output:** Top-$k$ neighbors
3:
4: $\hat{\mathbf{q}} \leftarrow$ QuantizeQuery($\mathbf{q}$)
5: // Stage 1: Coarse quantized search
6: $dist_{\text{quantized}}(\hat{\mathbf{q}}, \hat{\mathbf{x}}) = s_{\text{dist}} \cdot \sum_{j=1}^{d} (\hat{q}^{(j)} - \hat{x}^{(j)})^2$
7: $C \leftarrow \mathcal{G}.\text{Search}(\hat{\mathbf{q}}, ef = N_c \times m_{ef})$
8: // Stage 2: Asymmetric refinement
9: $dist_{\text{asym}}(\mathbf{q}, \tilde{\mathbf{x}}_{N_c}) = \sum_{j=1}^{d} \left(q^{(j)} - \tilde{x}^{(j)}\right)^2$
10: **for each** $\hat{\mathbf{x}}_i \in C$ **do**
11:     Reconstruct $\tilde{x}^{(j)} = \frac{\hat{x}^{(j)}}{\text{scale}_j} + \min_j$ and compute $d_i^a$
12: **end for**
13: $\mathcal{A} \leftarrow$ Sort by $d^a$
14: // Early termination
15: **if** $|\mathcal{A}| > k$ **then**
16:     Terminate early $\iff$ (gap > $\tau_{\text{gap}}$) ∨ (ratio > $\tau_{\text{ratio}}$)
17: **end if**
18: // Stage 3: Exact reranking
19: $d_{\text{exact}}(\mathbf{q}, \mathbf{x}) = \sum_{j=1}^{d} (q^{(j)} - x^{(j)})^2$
20: **for** $i = 1$ **to** $N_{\text{rerank}}$ **do**
21:     Retrieve $\mathbf{x}_i$ and compute $d_i^e$
22: **end for**
23: $\mathcal{R} \leftarrow \mathcal{R}_{\text{exact}} \cup$ remaining from $\mathcal{A}$
24: **return** Top-$k$ from sorted $\mathcal{R}$

## 3.3 Computational Acceleration via SIMD

To maximize computational efficiency on modern processors, AQR-HNSW incorporates architecture-specific Single Instruction Multiple Data (SIMD) optimizations that exploit data-level parallelism across the three-stage search pipeline described in Section 3.2.

For Stage 1's coarse search, the 8-bit quantized distance computation is vectorized by loading multiple uint8 dimensions from both query and database vectors into SIMD registers, computing absolute differences in parallel, then unpacking these 8-bit values to 16-bit representations before squaring to prevent overflow. The squared differences are accumulated using parallel multiply-add operations, with loop unrolling employed to hide memory latency.

Query encoding accelerates the float-to-uint8 conversion by vectorizing the quantization pipeline: multiple floating-point query dimensions are loaded simultaneously, followed by parallel subtraction of learned minimum values, multiplication by scaling factors, rounding to nearest integers, and clamping to the [0, 255] range to get all operations completed within SIMD registers.

Stage 3's exact distance computation leverages fused multiply-add (FMA) instructions, which compute $(q^{(j)} - x^{(j)})^2$ + sum as a single operation, reducing instruction count and register pressure while improving numerical precision.

The system performs compile-time architecture detection with automatic fallback: prioritizing AVX-512 when available, degrading gracefully to AVX2, SSE2, or NEON, and ultimately defaulting to scalar code on unsupported platforms. This ensures maximizing throughput on SIMD-capable processors without modifying the algorithmic framework.

## 4 Evaluation

### 4.1 Experimental Setup

All experiments were conducted on an Apple MacBook Pro with M4 Pro processor and 24GB unified memory. To isolate algorithmic contributions, Sections 4.1-4.5 present scalar baseline results, while Section 4.6 evaluates SIMD-accelerated performance. We evaluate AQR-HNSW across four standard benchmark datasets spanning diverse dimensions and sizes (Table 1).

To comprehensively evaluate performance and quantify the impact of each optimization, we compared the following HNSW algorithms: (1) *Baseline HNSW [23]*, serving as the unoptimized reference; (2) *Scalar Quantization*, applying baseline 8-bit quantization to HNSW; (3) *Density-aware HNSW (DHNSW)* [18], an enhanced HNSW incorporating local density estimation; and (4) our *AQR-HNSW*.

We measure performance using four key metrics. **Queries Per Second (QPS)** measures the throughput which indicates the number of nearest neighbor queries can be processed per second. **Recall@10** measures the accuracy as the fraction of true top-10 nearest neighbors correctly identified. **Build Time** captures the total time to construct the index graph. **Query latency** represents the performance for the queries processed during search phase.

### 4.2 Query Performance vs. Recall

Figure 2 presents the QPS-recall trade-off across four datasets. The results demonstrate distinct performance patterns that correlate with dataset dimensionality and structure.

AQR-HNSW: Accelerating Approximate Nearest Neighbor Search via Density-aware Quantization and Multi-stage Re-ranking

Table 1: Benchmark Dataset Characteristics

| Dataset | Vectors | Dimensions | Type | Size |
| --- | --- | --- | --- | --- |
| Fashion-MNIST [39] | 60,000 | 784 | Image | 183 MB |
| SIFT-1M [17] | 1,000,000 | 128 | SIFT descriptors | 488 MB |
| GIST-1M [17] | 1,000,000 | 960 | GIST features | 3.66 GB |
| GloVe-300D [29] | 100,000 | 300 | Word embeddings | 114 MB |

Table 2: Speedup Factor Comparison Across Datasets

| Method | Fashion-MNIST | SIFT-1M | GIST-1M | GloVe-300D |
| --- | --- | --- | --- | --- |
| Baseline HNSW | 1.00× | 1.00× | 1.00× | 1.00× |
| DHNSW | 1.12× | 1.18× | 1.39× | 1.12× |
| AQR-HNSW | 5.36× | 2.83× | 3.81× | 4.16× |

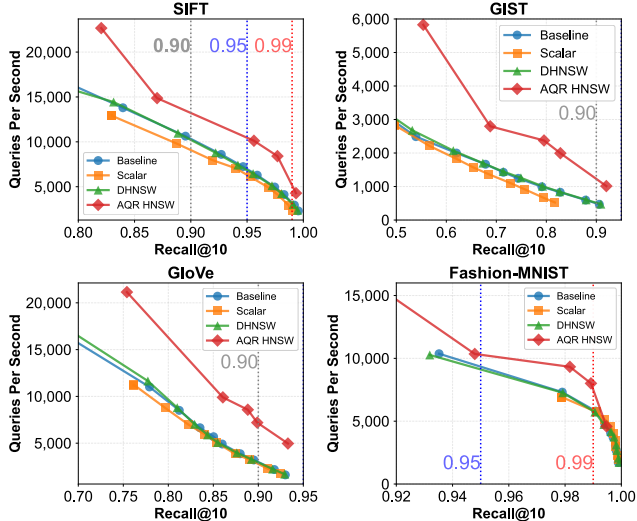

Figure 2: Recall@10 vs. QPS for different approaches — AQR-HNSW consistently achieves superior performance in the high-recall region (>90%).

**High-Dimensional Superiority:** On **GIST-1M**, AQR-HNSW sustains strong throughput even in the 0.9 recall regime where baseline HNSW degrades sharply. On **Fashion-MNIST**, AQR-HNSW has a high QPS while maintaining the recall at 0.99 as compared to baseline HNSW and DHNSW. AQR-HNSW's *density-aware quantization* preserves local ordering in dense neighborhoods, coupled with *early termination* that exploits larger high-D distance gaps to stop confidently without sacrificing recall.

**Medium-Dimensional Efficiency:** On **SIFT-1M**, AQR-HNSW gains 2–3 times high QPS and the recall approaches 0.95 as compared to baseline HNSW and DHNSW. **GloVe-300D** shows the similar trend of 2-2.5 times of QPS with a marginal recall loss compared to the baseline HNSW and DHNSW. The advantage comes from AQR-HNSW's multi-stage (*coarse-then-exact*) re-ranking design, avoiding full-precision work on every possible candidate.

### 4.3 Index Construction Efficiency

Table 2 shows that on million-point datasets (SIFT-1M and GIST-1M) AQR-HNSW achieves a 3–5× speedup over baseline HNSW and a 2-4× speedup over DHNSW. On smaller collections like Fashion-MNIST and GloVe-300D it exceeds a 4× reduction in build time (and higher recall in search phase). These gains come from density-aware quantization during construction using the sampling, assigns tighter per-dimension ranges where datapoints are clustered, and adjust proper construction parameters.

### 4.4 Sensitivity Analysis

Figure 3 reveals how AQR-HNSW hyperparameters affect the performance recall trade-offs, providing crucial insights for deployment optimization.

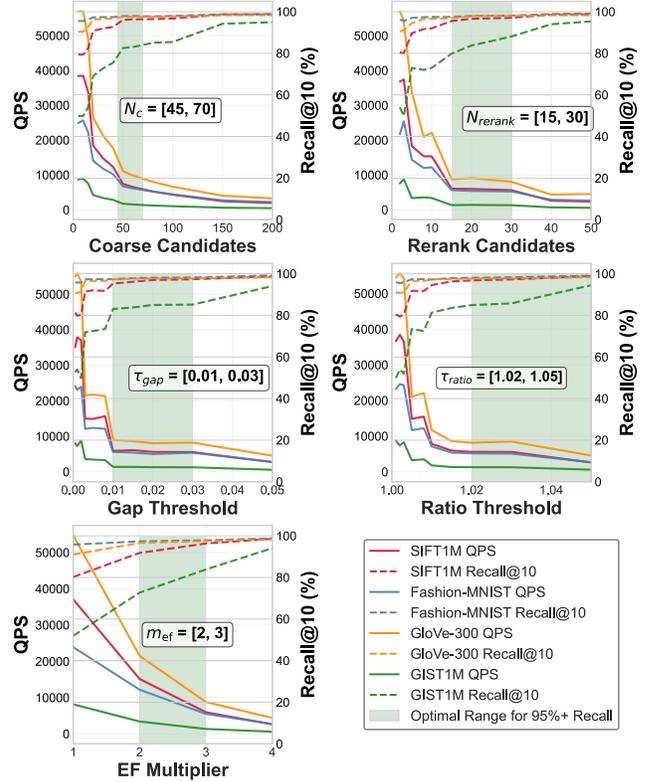

Figure 3: Ablation Study — Impact of hyper-parameters on QPS (solid lines) and Recall@10 (dashed lines).

Across the datasets, we find that tuning coarse candidates to about 45–70 maximizes recall improvements before they plateau and QPS falls linearly, while a rerank top-K of 15–30 delivers the best speed-accuracy trade-off (with higher values reserved for > 95% recall needs). Likewise, setting gap thresholds to 0.02–0.03 and ratio thresholds to 1.01–1.02 preserves high recall but triggers early termination for significant speedups, and using an EF multiplier of 2–3 offers optimal gains; further increases yield only marginal recall benefits at disproportionate QPS cost. In well-quantized spaces, true nearest neighbors exhibit distinct distance gaps and ratios, allowing conservative, moderate-depth search strategies to maintain accuracy with minimal overhead. The green region in Figure 3 highlights the range of each parameters that lead to high QPS with a consistent recall of 95% or higher. Based on extensive parameter sweeps, Table 3 summarizes the recommended configurations for different target recall levels from this study.

Table 3: Recommended Parameter Settings for Target Recall

| Target Recall | Coarse Cand. | Rerank K | Gap Thresh. | Ratio Thresh. | EF Mult. |
| --- | --- | --- | --- | --- | --- |
| 0.85–0.90 | 35 | 12 | 0.020 | 1.015 | 2 |
| 0.90–0.95 | 45 | 16 | 0.015 | 1.012 | 2 |
| 0.95–0.97 | 55 | 20 | 0.012 | 1.010 | 3 |
| 0.97+ | 70 | 28 | 0.010 | 1.008 | 3 |

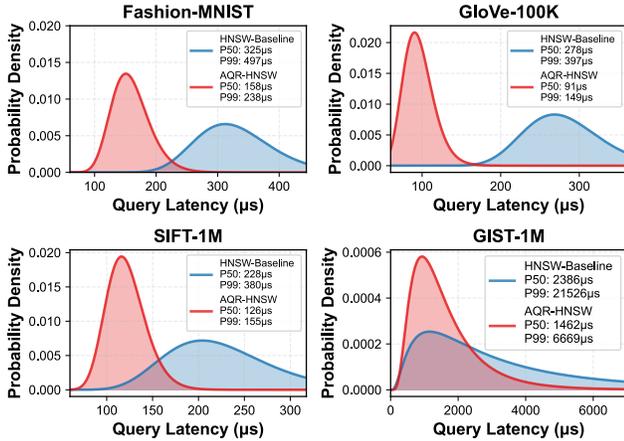

Figure 4: Query latency distribution for HNSW-Baseline and AQR-HNSW. DHNSW shows the same latency as baseline HNSW. AQR-HNSW consistently achieves better latency, showing significant improvements in both median (P50) and tail (P99) latencies.

### 4.5 Query Latency Consistency Analysis

Figure 4 demonstrates the query latency distribution, highlighting AQR-HNSW's consistent high performance over the search phase. We queried 10K samples on Fashion MNIST and Glove and 100k on SIFT and GIST, respectively. Queries processed by AQR-HNSW have dramatically denser probability density across all datasets, indicating more predictable performance. AQR-HNSW also mitigates the long tail latency via adaptive early termination and quantization-aware re-ranking, resulting in more stable latency performance under heavy query loads. Overall, AQR-HNSW achieves upto 3.23x lower P99 (on GIST) and 3x lower P50 (on GloVe) latencies as compared to baseline HNSW and DHNSW.

### 4.6 SIMD Enhancement for AQR-HNSW

To validate the computational benefits of the SIMD optimization layer described in Section 3.3, we evaluated AQR-HNSW with and without SIMD acceleration on two architectures: Intel Xeon Gold 6444Y (x86 with AVX-512 [14]) and Apple M4 Pro (ARM with NEON [2]).

Figure 5 shows that SIMD enhancement boosts both AQR-HNSW and baseline HNSW's performance at ISO-recall. Furthermore, AQR-HNSW benefits substantial QPS gains from SIMD acceleration: on x86 with AVX-512 [14], AQR-HNSW's throughput rises from baseline HNSW's 15,742 QPS to 47,364 QPS (3.0×) on SIFT1M dataset; and on ARM NEON [2], AQR-HNSW has a 3.2× higher QPS on GIST1M. Across the board, SIMD version of AQR-HNSW consistently 2.5–3.3× high QPS on different hardware platforms.

### 4.7 Comparison with Other ANN Approaches

Figure 6 presents a comprehensive QPS vs. Recall comparison against 12 leading ANN on the SIFT-1M benchmark which is analyzed in the following paragraph, benchmarking AQR-HNSW's performance in broader landscape of ANN search methods.

AQR-HNSW dominates the critical high-recall regime, delivering $2 - 2.5\times$ higher throughput than leading methods such as FAISS [19], NSG [37], and DiskANN [15]. This advantage stems from our *integrated* optimization strategy: density-aware quantization that minimizes approximation errors in pivotal regions, multi-stage early termination that aborts unnecessary searches, and asymmetric distance computation that accelerates exact re-ranking without sacrificing accuracy. In the speed-optimized zone (recall 0.4-0.7), quantization-only methods such as FAISS-PQ16 [16] and FAISS-SQ8 [19] achieve the highest throughput through aggressive compression but degrade rapidly as the recall increases. The balanced region (0.7-0.9) favors graph-based approaches: HNSW-M16 [23], NSG [37], and DiskANN [15] form a competitive cluster, while hybrid methods (FAISS-IVFPQ, FAISS-HNSWSQ) [19] underperform despite combining quantization with graph navigation. Tree-based methods (Annoy variants [4], FLANN [26]) and LSH-based FALCONN [1] show mediocre performance throughout. Most critically, in the accuracy-optimized region approaching perfect recall, AQR-HNSW maintains over 10,000 QPS at 95% recall while even strong competitors like HNSW-M16 [23] and DiskANN [15] drop below 5,000 QPS, demonstrating how its adaptive mechanisms specifically excel where traditional algorithms face exponential slowdowns.

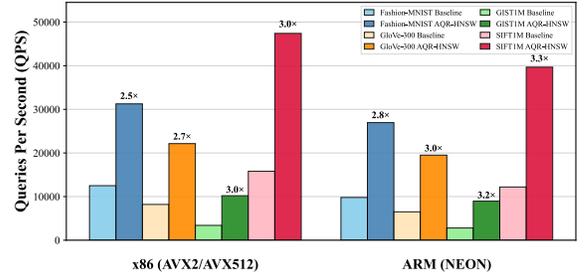

Figure 5: Performance comparison of AQR-HNSW against baseline HNSW with SIMD optimizations — The bars show QPS at 95% recall on x86 (AVX2/AVX512) [14] and ARM (NEON) [2] architectures, with average speedups of 2.8× and 3.1× respectively.

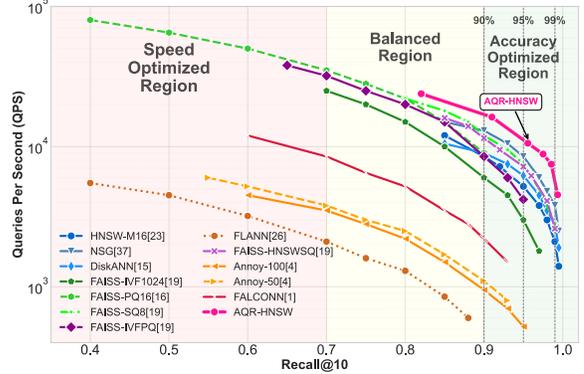

Figure 6: Comprehensive benchmark comparison of AQR-HNSW against state-of-the-art ANN algorithms on SIFT-1M dataset.All methods evaluated using scalar implementations (no SIMD) on Apple M4 Pro, single-threaded, to isolate algorithmic improvements. The plot is divided into three regions: speed-optimized ($< 90\%$ recall), balanced ($90-95\%$ recall), and accuracy-optimized ($> 95\%$ recall).

## 5 Conclusion

This paper presented AQR-HNSW, an accelerated graph-based similarity search framework via density-aware adaptive quantization and multi-stage reranking. AQR-HNSW jointly optimizes computation and accuracy by achieving 4× data compression while preserving neighborhood relationships, and reducing unnecessary exact computations by 35-60%. Our evaluation on billion-scale datasets demonstrates that AQR-HNSW can deliver 2.5-3.3× query throughput improvement across 80-99% recall targets, with optimal performance at 90-98% recall, which is critical for production deployments. Vector databases are the backbone of AI infrastructure, from trillion-scale RAG to billion-user recommendation engines, our density-aware design will become essential patterns for scaling.